\documentclass[pre,aps,twocolumn,floatfix]{revtex4}

\usepackage[dvips]{color,graphicx}
\usepackage{amsmath}
\usepackage{amsfonts}
\usepackage{bm}

\begin{document}

\date{\today}
\title{ A generalized Monte Carlo loop algorithm for frustrated Ising models }
\author{Yuan Wang}
\affiliation{Department of Applied Mathematics, University of Waterloo, Ontario, N2L 3G1, Canada} 
\author{Hans  De Sterck}
\affiliation{Department of Applied Mathematics, University of Waterloo, Ontario, N2L 3G1, Canada} 
\author{Roger G. Melko}
\affiliation{Department of Physics and Astronomy, University of Waterloo, Ontario, N2L 3G1, Canada}

\begin{abstract}

We introduce a Generalized Loop Move (GLM) update for Monte Carlo simulations of frustrated Ising models on two-dimensional lattices with bond-sharing plaquettes. 
The GLM updates are designed to enhance Monte Carlo sampling efficiency when the system's low-energy states consist of an extensive number of degenerate or near-degenerate spin configurations, separated by large energy barriers to single spin flips.  Through implementation on several frustrated Ising models, we demonstrate the effectiveness of the GLM updates in cases where both degenerate and near-degenerate sets of configurations are favored at low temperatures.  The GLM update's potential to be straightforwardly extended to different lattices and spin interactions allow it to be readily adopted on many other frustrated Ising models of physical relevance.
\end{abstract}
\maketitle

\section{Introduction}

Monte Carlo (MC) simulations are among the most ubiquitous computational tools used in statistical physics and material science.  Modern MC methods, evolved from the original Metropolis method of the 1950s \cite{metrop}, have become increasingly sophisticated with the advent of cluster moves, histogram reweighting methods, parallel tempering, and other technical advances \cite{Liu}.  This sophistication, coupled with the continuing increase in available computing power, allows MC methods to simulate classical and quantum statistical mechanical systems of a level of complexity unimaginable even a decade ago.  Indeed, the remarkable growth in the size of systems accessible to MC simulations is owed as much to the advancement of algorithm technology as it is to the increase in available raw CPU power through Moore's law \cite{moore}.

Lattice magnetic systems offer some of the greatest challenges to MC practitioners.  In fact, it was realized early on that MC simulations of the ferromagnetic Ising model employing simple local updates (single spin flip (SSF) algorithms) suffer severe critical slowing down -- a rapid increase in autocorrelation times -- near a second-order phase transition.  This stimulated work on non-local (cluster or collective-mode) algorithms, such as the well-known Swendsen-Wang \cite{SW} and Wolff \cite{Wolff} algorithms.  A different situation that also requires non-local updates is known to occur in broad classes of frustrated magnetic Ising models.  Such models, typically identified by predominantly antiferromagnetic (AFM) or random interactions, can have disordered groundstates which consist of an extensive number of equal-energy (degenerate) spin configurations - such is famously the case for the triangular lattice Ising model with AFM interactions \cite{Wannier,Houtappel}.  In such models, simple SSF updates have the tendency to bring the configuration out of the degenerate manifold of groundstates, hence costing an energy proportional to the Ising interaction.  At sufficiently low temperatures, the Metropolis algorithm will reject such moves, inhibiting {\it ergodicity}, which is the ability of the MC simulation to explore the entire degenerate manifold of states in reasonable (i.e.~non-exponential) computing time. 

This issue was recently brought to the forefront in the broad class of {\it ice} or {\it vertex} models \cite{BN}.  Of largest interest are the so-called {\it spin ice} models \cite{SpinIce} -- Ising models on a frustrated pyrochlore lattice, which are realized experimentally in some rare-earth titanate compounds.  Ideal spin ice Ising models promote a disordered degenerate groundstate.  However, it was found that weak perturbations, such as occur from long-range dipolar interactions, can energetically favor one or several configurations \cite{LRO}.  Exploration of different configurations contributing to the degenerate groundstate is thus a crucial task for MC simulations, and can only be achieved through loop updates \cite{loops1,loops2}.
The dynamics of loops in spin ice and related models are also intimately tied to important physical phenomena such as the Kasteleyn transition \cite{PeterKast}, and the concept of fluctuations between topologically ordered ground-state sectors \cite{Topo}.
In addition, loop dynamics in MC simulations have become a crucial part of our understanding of aspects of experimental dynamics in real spin ice materials \cite{loops1}.

Since the identification of loop algorithms is topical not only as a simulation technique but as a probe into the physics of materials, it is remarkable that generalized loop algorithms do not exist for broad classes of different frustrated models.  Rather, each system typically needs to have an algorithm designed based on the particular constraints of its groundstate manifold.  Such is the case for vertex models \cite{BN}, spin ice \cite{loops1,loops2} (and related corner-sharing triangle models such as the Ising kagome AFM), and recently the fully-frustrated honeycomb Ising model \cite{Shawn}.  In the present work, we address this shortcoming by introducing a generalized loop algorithm for a large class of frustrated Ising models with bond-sharing plaquettes.

The paper is organized as follows.  In Section \ref{SectII}, we outline a Generalized Loop Move (GLM) update in detail for a class of two-dimensional frustrated Ising models, illustrating the specific examples of the triangular lattice Ising AFM, and the fully-frustrated square and honeycomb lattice Ising models.  We prove rigorously that the algorithm obeys detailed balance (Appendix \ref{appendixA}), and demonstrate explicitly that it reproduces the results expected in traditional MC simulations of these models (Section \ref{unpert}).  In Section \ref{pert}, we simulate extensions of the models, perturbed by weakening one bond per plaquette (an interaction that can arise experimentally in frustrated magnets when uniaxial pressure is applied \cite{Topo}).  In this case, the GLM update vastly outperforms traditional single-spin flips in the exploration of the low temperature physics of the models, in particular in finding a unique ordered groundstate.   
As discussed in Section \ref{discuss}, GLM update will help open the way for the efficient simulation of a wide range of frustrated magnetic models in the future.

\section{Generalized loop algorithm for Ising models.} \label{SectII}

\subsection{Ising models}

\begin{figure}
\includegraphics[width=8cm]{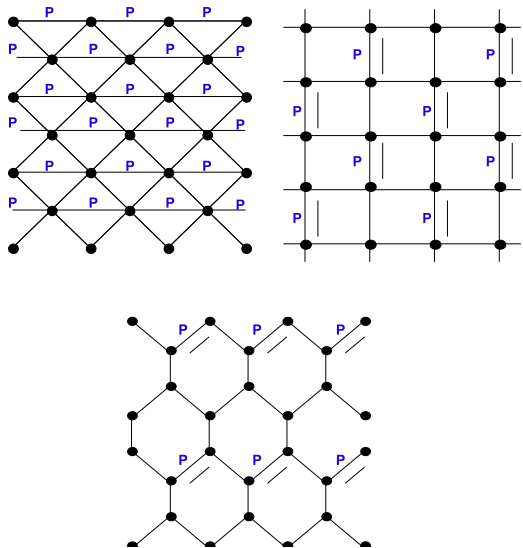}
\caption{Frustrated Ising spin models: AFM triangle (top left), FF square (top right) and FF honeycomb (bottom). The single bonds represent AFM bonds, the double bonds represent FM bonds, and the perturbed (weakened) bonds in the perturbed Ising models are indicated by ``P" (see Section \ref{pert}).}
\label{fig:Perturbed Fully Frustrated Ising Models}
\end{figure}

We consider classical Ising models that have Hamiltonians of the general form
\begin{eqnarray}
H=  \sum_{\left< i,j \right>} J_{ij} \delta_{ij}S^z_i S^z_j, \label{HAM}
   \end{eqnarray}
where spin variables $S^z_i$ can take the value of $\pm 1/2$.  
In this paper, we refer to \emph{unperturbed} models as those with $J_{ij} $ equal to a constant $J>0$, while \emph{perturbed} models (discussed in Section \ref{pert}) have some of the $J_{ij} = J' \neq J$ (but still positive).
The bond variables $\delta_{ij}$ are defined to have  a value of either $+1$ (``AFM") or $-1$ (``FM") for each bond on the lattice. In a given spin configuration, a bond is referred to as \emph{satisfied} if $\delta_{ij}S_{i}^{z}S_{j}^{z} < 0$ and \emph{unsatisfied} otherwise. An Ising model is referred to as \emph{frustrated} if it is impossible for any spin configuration to satisfy all of the bonds on the lattice.  
We consider both perturbed and unperturbed versions of the following two-dimensional (2D) periodic lattice models in this paper (Fig \ref{fig:Perturbed Fully Frustrated Ising Models}):
\begin{enumerate}
\item The triangular lattice AFM, where all $\delta_{ij} = 1$.
\item The fully frustrated square lattice.
\item The fully frustrated honeycomb lattice.
\end{enumerate}
For the latter two lattices, uniform AFM interactions are unfrustrated; full frustration can be induced by enforcing the constraint that the product of the sign of the bond variables around {\it each} closed plaquette (square or hexagon) satisfies
\begin{eqnarray}
\prod_{ij \in {\rm plaq}} \delta_{ij} = -1, \label{gauge}
\end{eqnarray}
e.g. $\delta_{ij}=-1$ for one bond, and $\delta_{ij}=1$ for the rest.  Since each plaquette is frustrated such interactions are called {\it fully-frustrated} (FF).  Thus defined, the three frustrated Ising models that we are considering are known to have disordered groundstates with an extensive number of equal-energy configurations \cite{Wannier,Houtappel,square1,Moessner,WolffZ}.  The geometry of the FM and AFM bonds for each lattice is illustrated in Fig.~\ref{fig:Perturbed Fully Frustrated Ising Models}.

\subsection{Single-spin flips}

It is common to study the finite temperature ($T$) and groundstate properties of such models using Markov Chain Metropolis Monte Carlo methods.
Traditionally, single spin flip (SSF) updates are employed in the Monte Carlo algorithm;
one simply attempts to flip a single spin (one at a time) by computing the corresponding change in energy ($\Delta E$) 
and then accepting the update with a Metropolis condition using probability:
\begin{equation}
P_{flip}=min(1,\: e^{-\frac{\Delta E}{T}}) \label{eq:SSF Pflip}
\end{equation}
In frustrated Ising models, many spin configurations are local
energy minima, and most SSF attempts cost a large energy proportional to $J$.
Based on Eq.~(\ref{eq:SSF Pflip}), the probability
of accepting such a SSF update decreases exponentially
as the temperature decreases. Consequently, at $T \ll J$, once a SSF simulation has reached
one of the local energy minima it is very unlikely for the simulation
to accept any SSF updates. As a result,
SSF updates tends to become ``frozen'' into
one of the local energy minima at $T \ll J$ (i.e.~ergodicity is lost).

One can observe that in these local energy minima, there are often groups
of spins that could be flipped together without changing the number
of unsatisfied bonds (and hence without significantly changing the
energy of the system).  Cluster algorithms take advantage of
this observation to find and flip such groups of spins \cite{BN}.

\subsection{A Generalized Loop Move}

Motivated by the above, we introduce a Generalized Loop Move Algorithm (GLM), designed to improve the sampling of equal (or nearly-equal) energy configurations near the groundstates of 2D periodic frustrated Ising models with bond sharing plaquettes.  Like previous algorithms designed to work on specific Ising models \cite{BN,LRO, Shawn}, the algorithm works by identifying groups or clusters of spins that could be flipped together without changing the number of unsatisfied bonds, and then attempting to flip them according to a standard Metropolis update.  In this way, no energy costs (proportional to $J$) are accrued, which as mentioned above can inhibit simulation dynamics at low temperatures.

To find these groups of spins, the GLM algorithm uses the concepts of dual graphs and matchings from graph theory. In this paper, dual graphs are defined in the usual graph theory sense. The dual nodes represent the faces/plaquettes in the original (``primal") spin lattice, and the dual edges connect dual nodes which correspond to faces that share a bond in the primal spin system.  See Fig.~\ref{fig:IsingLatticesandDuals} for how the dual graphs are defined on the triangle, square and honeycomb lattices respectively. If we take a spin system and examine its dual graph, we can consider each dual edge to be inside a \emph{pseudo-matching} whenever the corresponding bond in the primal spin lattice is unsatisfied. This pseudo-matching is not a matching in the rigorous graph theory sense. In a graph theory matching, each node can only be adjacent to at most one edge in the matching. On the other hand, in the pseudo-matching described in this paper, each dual node can be adjacent to multiple dual edges in the pseudo-matching. There are a few interesting properties to note about this pseudo-matching:
\begin{enumerate}
\item In any pseudo-matching that corresponds to a valid spin configuration, each dual node is adjacent to at least one dual edge in the pseudo-matching because each plaquette contains at least one unsatisfied bond in a fully frustrated Ising model.
\item Each plaquette in a fully frustrated Ising model always has an odd number of unsatisfied bonds. Consequently in any pseudo-matching that correspond to a valid spin configuration, each dual node is adjacent to an odd number of dual edges in the pseudo-matching.
\item Each valid ground-state spin configuration and its spin inverse corresponds to a unique pseudo-matching.
\item There is a bijective correspondence between pairs of valid spin configurations and their spin inverses with the minimal number of unsatisfied bonds (one per plaquette) and perfect matchings in the dual graph. A perfect matching is a matching where each node is adjacent to exactly one edge in the matching.
\end{enumerate}
Given a spin configuration, the problem of finding another spin configuration with the same number of unsatisfied bonds is equivalent to finding a pseudo-matching in the dual graph with the same number of dual edges as the pseudo-matching corresponding to the original spin configuration. In graph theory, a basic strategy for finding a new matching, 
with the same number of edges as an existing matching,
is to find an alternating cycle. 
Given some matching, an alternating cycle consists of edges that are alternatingly inside and outside the matching. For each edge in the alternating cycle, we can add it to the matching if it was not originally in the matching, and remove it from the matching otherwise, to get another matching of with the same number of edges. This operation is known as taking a \emph{symmetric difference} between the matching and the alternating cycle.
The same strategy can be used to find a new pseudo-matching with the same number of dual edges as an existing pseudo-matching. Thus, if we start with a spin configuration and want to find another with the same number of unsatisfied edges, we can consider its corresponding pseudo-matching, use an alternating cycle to find another pseudo-matching, and then map this back to a spin configuration.
If we look at this series of actions from the perspective of the spin system, we have found a group of spins delineated by the alternating cycle in the dual graph that can be flipped without changing the number of unsatisfied bonds.

The motivation for looking at the dual graph of the spin system comes from the work by Bieche {\it et.~al.}~\cite{Bieche}. In that work, the authors devised an algorithm that uses a modified dual graph of a spin system to find its groundstates. The algorithm takes the dual graph of the spin system and modifies it so that there is a bijective correspondence between perfect matchings in the modified dual graph and valid spin configurations. Weights are applied to each edge in the modified dual graph so that the weight of a perfect matching is equal to the energy of the corresponding spin configuration. Consequently, the problem of finding a groundstate in the spin system is transformed into the problem of finding a minimally weighted perfect matching in the dual graph, for which a polynomial time algorithm has been devised by Edmonds {\it et.~al.}~\cite{edmonds1,edmonds2}.  One anticipates that the GLM update, designed in this paper to operate on fully frustrated Ising models, could be extended to work on a broader class of frustrated Ising models by adapting it to use the modified dual graph described in Bieche's paper.

\subsection{Description of the Algorithm}

\begin{figure}
\includegraphics[width=8cm]{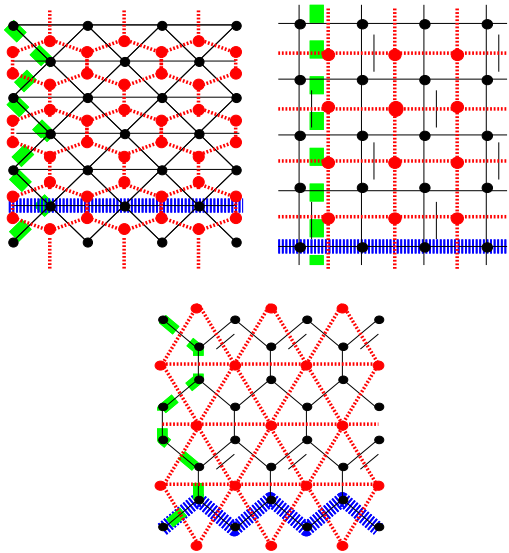}
\caption{Ising spin models and the corresponding dual graphs: AFM triangle (top left), FF square (top right), and FF honeycomb (bottom). The solid black lines represent primal spin lattice and the thick dashed red lines represent the dual lattice. The dashed shading of primal lattice along the vertical axis represent the \emph{Vertical Reference Cut} and the hashed shading of primal lattice along the horizontal axis represent the \emph{Horizontal Reference Cut}.}
\label{fig:IsingLatticesandDuals}
\end{figure}

Like previous loop and cluster algorithms, 
The Generalized Loop Move (GLM) is designed to supplement single spin flip (SSF) updates.  
Each iteration of the whole algorithm involves first attempting
single spin flips on each spin in the lattice, and then attempting
a fixed number of GLMs on the whole lattice. In this paper, we have chosen to perform $N/30$ GLM attempts after each SSF sweep (where $N$ is the number of spins in the lattice). In
each GLM attempt, we try to identify a group of
spins that can be flipped together without changing the number of unsatisfied
bonds in the lattice, compute the probability with which we should
accept the flip so as to maintain detailed balance, and then accept
the flip according to this probability.

For ease of description, we divide the GLM algorithm
into 3 subtasks: the Partition Finding subtask, the Loop Generating
subtask and the Acceptance Probability Calculating subtask. The Partition
Finding subtask describes how the algorithm finds a group of spins
that can be flipped together without changing the number of unsatisfied
bonds. The Loop Generating subtask describes how the algorithm finds
cycles in the dual graph that are used by the Partition Finding subtask
to identify the above-mentioned group of spins. Lastly, the Acceptance
Probability Calculating subtask, as the name implies, describes how the algorithm calculates
the acceptance probability with which we should accept suggested flips
to maintain detailed balance.

\begin{figure}
\includegraphics[width=8cm]{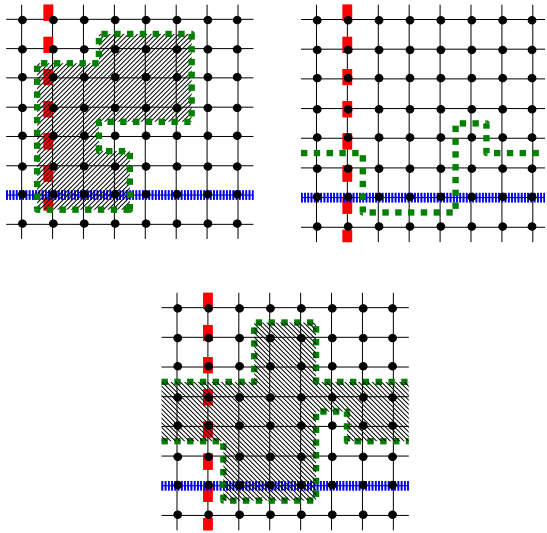}
\caption{Types of cycles in the dual graph: closed loop (top left), cut (top right), complementary union of cuts (bottom). The dashed shading on the primal lattice along the vertical axis represents the \emph{Vertical Reference Cut} and the hashed shading of primal lattice along the horizontal axis represents the \emph{Horizontal Reference Cut}. The dashed green line on the dual graph represents the set of dual edges making up the cycle. The hash shaded area represents the spins enclosed by the Closed Loop or Complementary Union of Cuts.}
\label{fig:LoopTypes}
\end{figure}

To describe the algorithm precisely, we will first establish the following
definitions and notations:
\begin{enumerate}
\item A bond between spin $i$ and spin $j$ is referred to as $B_{ij}$ and its
value is denoted as $J_{ij}\delta_{ij}$ where $J_{ij}$ and $\delta_{ij}$ are defined as in Eq.~(\ref{HAM}); its dual is referred to as $\widehat{B_{ij}}$.
\item A set of dual edges is called a {\it partition boundary} if deleting all
the bonds dual to these dual edges would divide the original lattice
into two connected components.
\item A {\it Horizontal Reference Cut} refers to a specific set of bonds that
wraps horizontally around the lattice; a {\it Vertical Reference Cut}
similarly refers to a specific set of bonds that wraps vertically
around the lattice. These reference cuts are used in the algorithm
to determine whether a set of dual edges constitutes a partition boundary.
The set of bonds chosen is arbitrary.
See Fig.~\ref{fig:IsingLatticesandDuals} for how the Horizontal and Vertical
Reference Cuts are defined in this paper on the AFM triangle, FF square and FF honeycomb lattices
respectively. Other choices of horizontal and vertical reference
cuts that wrap horizontally or vertically around the lattice would
work just as well.
\item A cycle is called a {\it simple cycle} if it does not intersect itself. If a simple cycle of dual edges in the dual lattice is dual to an
even number of bonds in the Horizontal Reference Cut and an even number
of bonds in the Vertical Reference Cut, then we refer to the cycle
of dual edges as a {\it closed loop}. See Fig. \ref{fig:LoopTypes} for
an example of a closed loop. It can be shown that such a cycle of
dual edges forms a partition boundary.
\item If a simple cycle of dual edges in the dual lattice is dual to an
odd number of bonds in the Horizontal Reference Cut or an odd number
of bonds in the Vertical Reference Cut, then we refer to the cycle
of dual edges as a {\it cut}. See Fig. \ref{fig:LoopTypes} for an example of
a cut. It can be shown that such a cycle of dual edges does not form
a partition boundary.
\item If the union of two non-intersecting cuts is dual to an even number
of bonds in the Horizontal Reference Cut and an even number of bonds
in the Vertical Reference Cut, then it is called a {\it complementary union
of cuts}. See Fig.~\ref{fig:LoopTypes} for an example
of a complementary union of cuts. It can be shown that such a union
forms a partition boundary. 
\end{enumerate}

\subsection{Partition Finding Subtask\label{sub:Partition-Finding-Subtask}}

In the Partition Finding subtask, we try to find a partition boundary
that is either a closed loop or a complementary union of cuts. Once
we find such a partition boundary, we can use the Acceptance Probability
Calculating subtask (Section \ref{sub:Acceptance-Probability-Calculating}) to
compute the probability with which we should flip all the spins in
the smaller of the two partitions formed by the partition boundary
so as to maintain detailed balance. Then we flip all the spins inside
the smaller of the two partitions with the probability just computed.
When all the spins in either partition are flipped, all the satisfied
bonds dual to the partition boundary become unsatisfied, all the
unsatisfied bonds dual to the partition boundary become satisfied
and all other bonds remain unchanged. To ensure that the number of
unsatisfied bonds remains unchanged when we flip all the spins inside
the partition we want to choose a partition boundary that is dual
to an equal number of satisfied and unsatisfied bonds. One way to
satisfy this requirement is to choose a partition boundary that is
alternatingly dual to satisfied and unsatisfied bonds. To find such
a partition boundary, we use the Loop Generating subtask (Section \ref{sub:Loop-Generating-Subtask})
to find a simple cycle in the dual lattice that is alternating dual to satisfied
and unsatisfied bonds. This simple cycle may be a closed loop or a cut. If
it is a closed loop, it forms a partition boundary and we have found
the desired partition. If it is a cut, then we use the Loop Generating
subtask to find a second cycle in the dual lattice. If the second
cycle is another cut that does not intersect the first one and the
union of the two cycles form a complementary union of cuts, then the
union of two cuts forms a partition boundary and we have found the
desired partition. If the second cycle does not meet the above requirements,
then we abort the Generalized Loop Move. The reason why we want to find complementary unions of cuts is that in perturbed models at $T
\ll J$, numerical results have shown that GLM using closed loops alone has a high probability of becoming frozen in a near-groundstate. On the other hand, GLM using a combination of closed loops and union of cuts can efficiently sample from the near-groundstates. It should be noted that when the algorithm finds a cut and then a closed loop, it is possible to discard the cut and keep the closed loop as a partition boundary.

\subsection{Loop Generating Subtask\label{sub:Loop-Generating-Subtask}}

The Loop Generating subtask is designed to find simple cycles of dual edges
that are alternatingly dual to satisfied and unsatisfied bonds for
the Partition Finding subtask to use. The main idea is that we start
at a random dual node in the dual lattice, choose in a random but
weighted way one of the dual edges adjacent to the dual node, go to
the other dual node adjacent to the dual edge, choose in a random
but weighted way a dual edge adjacent to this dual node (we will choose
among dual edges dual to satisfied bonds if the last dual edge is
dual to an unsatisfied bond and vice versa) and continue this way
until we revisit one of the dual nodes we have passed through earlier
(i.e. until the string of dual edges ends in a simple cycle). This simple cycle
is used by the Partition Finding subtask as described in Section \ref{sub:Partition-Finding-Subtask}.
More precisely, the algorithm can be described as follows:
\begin{enumerate}
\item Randomly choose a dual node. Call it $dualnode1$ and the set of dual
edges adjacent to it $\widehat{E}$.
\item Randomly select one of the dual edges $\widehat{B_{ij}}$ in $\widehat{E}$. The selection should be completely random for unperturbed models or based on some weighting scheme that favors strong unsatisfied bonds
and weak satisfied bonds for perturbed models. When the spins inside the partition are
flipped, the bonds dual to the partition boundary are made satisfied if they
were previously unsatisfied and unsatisfied if they were previously
satisfied. So favoring strong unsatisfied bonds and weak satisfied
bonds for inclusion in the cycle means favoring partitions that would
decrease the energy of the system if flipped. This increases the likelihood
of finding a partition that is energetically favored to be flipped
at low temperatures. Numerical results show that a weighting scheme is particularly important for perturbed triangle and square lattices where a GLM algorithm which selects partition boundaries completely at random tends to fail to find partitions that are energetically favored to be flipped in a reasonable amount of time. The weighting scheme chosen by the authors of this paper is:
\begin{widetext}
\begin{equation}
W(\widehat{B_{ij}})=\begin{cases}
e^{\alpha \beta[|J_{ij}|-mean_{\widehat{B_{ij}}\in\widehat{E}}(\{|J_{ij}|\})]} & \text{if \ensuremath{B_{ij}} is unsatisfied},\\
e^{-\alpha \beta[|J_{ij}|-mean_{\widehat{B_{ij}}\in\widehat{E}}(\{|J_{ij}|\})]} & \text{if \ensuremath{B_{ij}} is satisfied,}\end{cases}\label{eq:Weighting Scheme}\end{equation}
\end{widetext}

where $\beta={1}/{T}$ and $\alpha$ is an arbitrary parameter
that affects how strongly the weighting scheme favors weak satisfied
bonds and strong unsatisfied bonds
\footnote{Note that in unperturbed models, $W(\widehat{B_{ij}})=1,\;\forall\widehat{B_{ij}}\in\widehat{E}$}.
 The authors chose $\alpha$ to
be 5. Preliminary numerical results showed that an $\alpha$ of 1 did not sufficiently favor partitions that were energetically favored to be flipped. As a result, GLM with an $\alpha$ of 1 showed slow equilibration of thermodynamic data near the groundstate. Consequently, an $\alpha$ of 5 was tried and the thermodynamic data showed fast equilibration. 

\item Add this dual edge to our chain of dual edges, go to the other dual
node adjacent to this dual edge and repeat the following steps until
we revisit a dual node we have already visited:

\begin{enumerate}
\item Call the set of dual edges adjacent to the current dual node $\widehat{E}$
\item Remove from $\widehat{E}$ all satisfied dual edges if the last dual
edge added to our chain is satisfied; otherwise, remove all unsatisfied
dual edges. 
\item Remove from $\widehat{E}$ all dual edges that would result in a
simple cycle with an odd number of dual edges if selected \footnote{
Note that (b) and (c) ensure that the resulting cycle/loop is alternatingly
dual to satisfied and unsatisfied bonds.}.
\item If $\widehat{E}$ is empty, then abort the Generalized Loop Move algorithm
and leave the spin configuration unchanged.
\item Randomly select a dual edge from $\widehat{E}$ based on a weighting
scheme that favors dual edges dual to strong unsatisfied bonds or
weak satisfied bonds. The author of this paper chose to use the weighting
scheme defined by Eq.~(\ref{eq:Weighting Scheme}).
\item Add the dual edge to the chain and go to the dual node on the other
end of the dual edge.
\end{enumerate}
\item Once we revisit a node, the simple cycle at the end of our chain of dual edges
is the desired simple cycle that we will use in the Partition Finding subtask.
\end{enumerate}

The computation time required to find a simple cycle through the algorithm described above is an important factor in the efficiency of the GLM algorithm. Numerical results concerning the amount of computation time required are discussed in Section \ref{sub:Simulation-Results}.

\subsection{Acceptance Probability Calculating Subtask\label{sub:Acceptance-Probability-Calculating}}

\begin{figure}[ht]
\includegraphics[width=8cm]{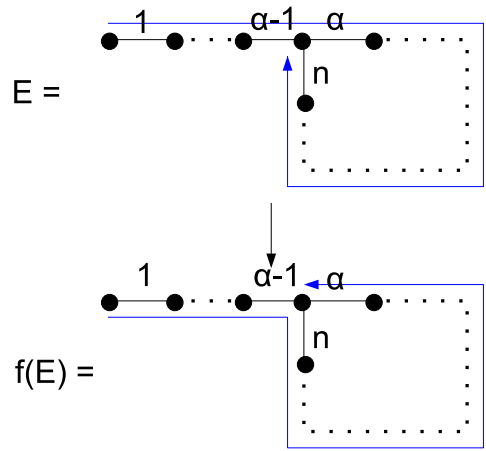}
\caption{The function $f$ described in the text.}
\label{fig:Function f}
\end{figure}

Once the Partition Finding subtask generates a partition boundary
in the dual graph, the Acceptance Probability Calculating subtask
computes an acceptance probability with which the GLM 
algorithm should flip all the spins inside the smaller partition
so as to maintain detailed balance. For ease of exposition, it is
convenient to establish the following definitions and notations:
\begin{itemize}
\item Define a function $f$ on the set of dual edge chains that end in
a simple cycle as follows: If $C=[e_{1\:}e_{2\:}...\: e_{\alpha}\: e_{\alpha+1}\:...\; e_{n-1}\: e_{n}]$
is a chain of dual edges that ends in a simple cycle $L=[e_{\alpha}\: e_{\alpha+1}\:...\; e_{n-1}\: e_{n}]$,
then $f(C)=[e_{1\:}e_{2\:}...\: e_{n}\: e_{n-1}\;...\; e_{\alpha+1}\: e_{\alpha}]$.
Essentially, $f$ takes a chain of dual edges that end in a simple cycle and
reverses the order of dual edges that are part of the simple cycle (see Fig.
\ref{fig:Function f}). Note: $f(f(C))=C$.
\end{itemize}

\begin{itemize}
\item Let $P_{select}(C,\: S)$ be the probability of the Loop Generating
subtask selecting the dual edge chain $C$ under the spin configuration
$S$. 
\item Denote the current spin configuration as $S_{1}$ and the spin configuration
after flipping all the spins inside the smaller partition as $S_{2}$. 
\end{itemize}
Then the acceptance probability is defined by the following expressions:
\begin{widetext}
\begin{itemize}
\item If the Partition Finding subtask found a partition boundary that is
a closed loop $L$ from a chain $C$, then:\begin{equation}
P_{accept}(C,\: S_{1})=min \left({ 1,\: e^{-\beta[E(S_{2})-E(S_{1})]}\frac{P_{select}(f(C),\: S_{2})}{P_{select}(C,\: S_{1})} }\right) \label{eq:Paccept 1}\end{equation}

In the unperturbed case where $E(S_{1})=E(S_{2})$, the expression
simplifies to: \begin{equation}
P_{accept}(C,\: S_{1})=min \left({ 1,\:\frac{P_{select}(f(C),\: S_{2})}{P_{select}(C,\: S_{1})} }\right) \end{equation}

\item If the Partition Finding subtask found a partition boundary that is
a complementary union of cuts $L_{1}\cup L_{2}$ from chains $C_{1}$and
$C_{2}$, then:\begin{equation}
P_{accept}((C_{1},C_{2}),\: S_{1})=min \left({ 1,\: e^{-\beta[E(S_{2})-E(S_{1})]}\frac{P_{select}(f(C_{1}),\: S_{2})P_{select}(f(C_{2}),\: S_{2})}{P_{select}(C_{1},\: S_{1})P_{select}(C_{2},\: S_{1})} }\right) \label{eq:Paccept2}\end{equation}

In the unperturbed case where $E(S_{1})=E(S_{2})$, the expression
simplifies to:\begin{equation}
P_{accept}((E_{1},E_{2}),\: S_{1})=min \left({ 1,\:\frac{P_{select}(f(C_{1}),\: S_{2})P_{select}(f(C_{2}),\: S_{2})}{P_{select}(C_{1},\: S_{1})P_{select}(C_{2},\: S_{1})} }\right)\end{equation}

\end{itemize}
\end{widetext}

The authors of this paper have chosen to compute $P_{select}(C,S_{1})$ and $P_{select}(f(C),S_{2})$ after a suitable partition boundary has been found.

A rigorous proof of detailed balance is included in Appendix \ref{appendixA}.

\section{Simulation results\label{sub:Simulation-Results} }

In this section we examine some prototypical data from Monte Carlo simulations of three frustrated Ising models, using simulations with SSF and GLM updates.  In section \ref{unpert}, thermodynamic data for unperturbed Ising models is compared between simulations employing only SSF, and simulations using both SSF and GLM updates.  Some GLM loop properties are characterized.  In section \ref{pert}, we examine the performance of the GLM updates on the three perturbed Ising models, which are constructed to have a finite-$T$ phase transition to a ground state with a unique spin configuration.  We demonstrate that GLM updates helps the simulation find the correct groundstate in cases where SSF updates fail.

\subsection{Results on unperturbed models} \label{unpert}

Metropolis Monte Carlo simulations are performed on the three aforementioned Ising models: AFM triangular, FF square and FF honeycomb (Fig.~\ref{fig:Perturbed Fully Frustrated Ising Models}).  A typical run employs an \emph{annealing} technique where the simulation is begun at a 
high temperature ($T$ = 0.5), and $T$ is gradually lowered through small steps until the system settled in its disordered degenerate groundstate. 
At each temperature, a fixed number of iterations of either SSF or a combination of SSF and GLM is applied and measurement estimators such as energy and magnetization are collected.  Each annealing simulation is performed twice: once with SSF only, and once with a combination of SSF and GLM updates. 

\begin{figure}
\includegraphics[width=6.5cm]{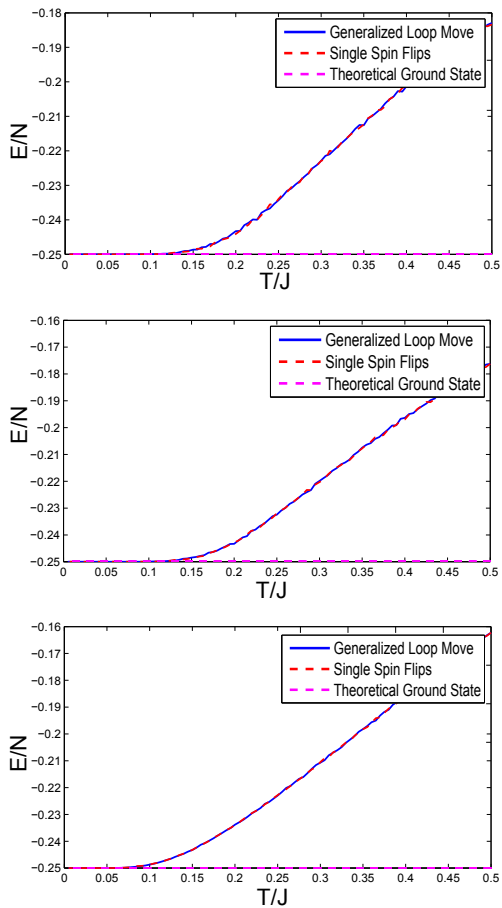}
\caption{Energy per spin (E/N) for the AFM triangular lattice (top), FF square lattice (middle), and FF honeycomb lattice (bottom). The theoretical true groundstate energy per spin for each system is -0.25}
\label{fig:AnnealingEnergy}
\end{figure}

The first numerical result we consider is the internal energy of the Ising models under an annealing simulation. 
As illustrated in Fig.~\ref{fig:AnnealingEnergy}, the internal energy measured as a function of temperature appears identical between SSF and GLM simulations, to within statistical error. This lends practical proof to the fact that the GLM updates are \emph{well behaved} (do not disrupt detailed balance in the Monte Carlo procedure), and find the same degenerate manifold of states as the SSF updates alone.
In addition to annealing simulations, we also tested the performance of SSF and GLM in \emph{quenched simulations} where the  
entire simulation (including the warm-up or equilibriation) is run at a single temperature.  We ran these single simulations at $T = 0.05$, 
finding that for the case of the AFM triangular and FF square lattice Ising models,
both SSF and GLM were able to find a groundstate configuration, the efficiency gain through GLM not being significant in this case.
However, for the FF honeycomb lattice, SSF simulations tend to be noticeably slow in finding the disordered groundstate, a situation remedied by employing GLM updates. This difficulty for SSF alone to simulate the low-temperature thermodynamic behavior of the FF honeycomb lattice was also noted in the work by Andrews {\it et.~al.}~\cite{Shawn} where a cluster update specific to the FF honeycomb was proposed. In Andrew's work, it was shown that in an annealing simulation, SSF alone tends to become frozen in one of the groundstates at low temperatures which is evident by the inability of the average magnetization to converge to zero (Fig. 3(c) of Ref.~\cite{Shawn}). On the other hand, both the algorithm proposed in that paper and GLM can find the correct average magnetization through annealing or quenched simulations.

\begin{figure}
\includegraphics[width=6.5cm]{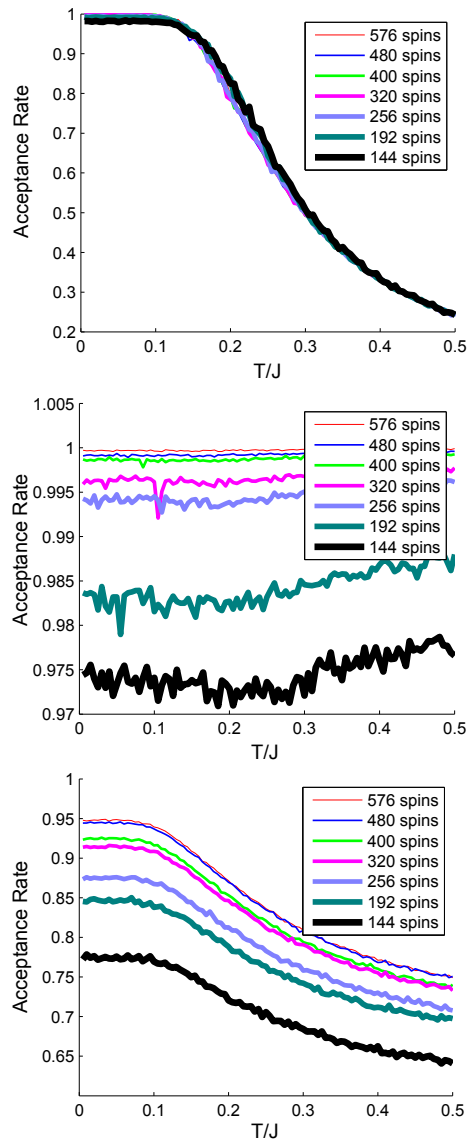}
\caption{Acceptance Rate of GLM on the AFM triangular lattice (top), FF square lattice (middle), and FF honeycomb lattice (bottom).}
\label{fig:AcceptanceRate}
\end{figure}

A meaningful characterization of the low-temperature dynamics of the GLM updates can be determined via the study of the acceptance rate of the algorithm.
We define the acceptance rate as the percentage of GLM attempts that result in {\it successful} group spin flips. The results are shown in Fig.~\ref{fig:AcceptanceRate}. For the AFM triangular lattice, the acceptance rate increases as temperature decreases and appears independent of lattice size. For the FF square lattice, the acceptance rate is high across the whole temperature range and appears to approach 1 as lattice size increases. For the FF honeycomb lattice, the acceptance rate is high across the temperature range, increasing as the temperature decreases, and apparently increasing with lattice size.  Most importantly, the fact that the acceptance rate is large (of order unity) in all cases, particularly for $T /J \leq 0.1$, suggests that the GLM updates are successful in exploring the manifold of degenerate configurations that contribute to the disordered groundstate.  This is in direct contrast to an algorithm employing SSF updates only, which are known to have exponentially suppressed acceptance rates for $T \ll J$.
 
\begin{figure}
\includegraphics[width=6.5cm]{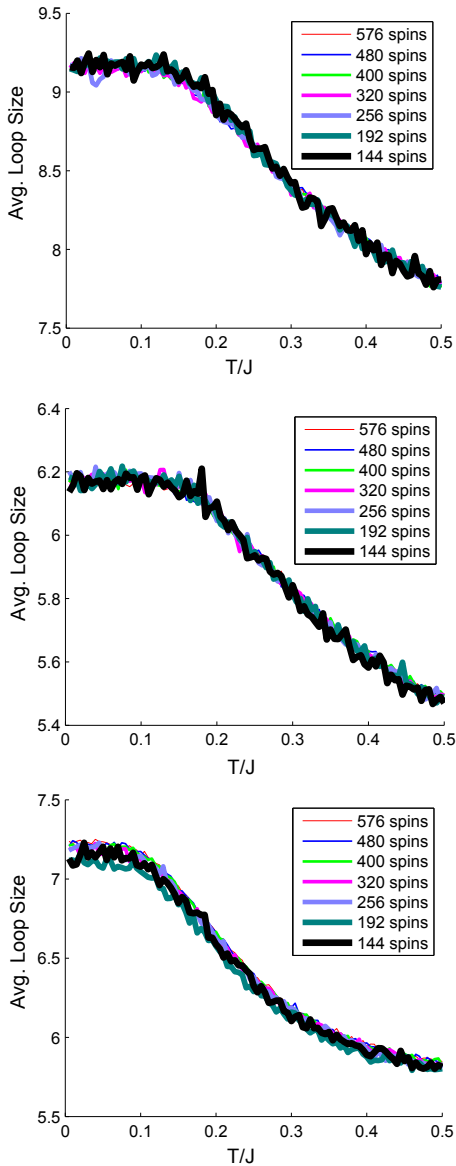}
\caption{Loop Size for the AFM triangular lattice (top), FF square lattice (middle), and FF honeycomb lattice (bottom).}
\label{fig:LoopSize}
\end{figure}

\begin{figure}
\includegraphics[width=6.5cm]{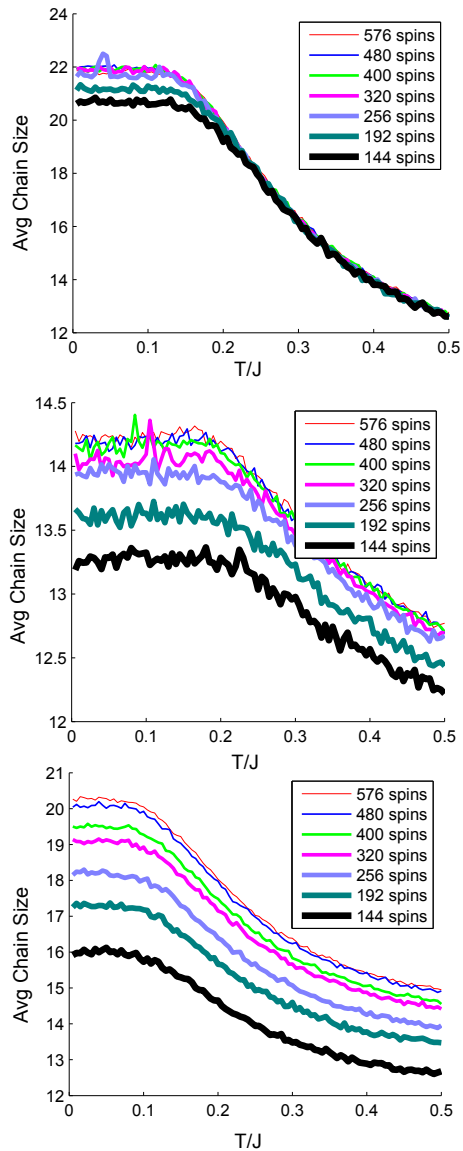}
\caption{Chain Size for the AFM triangular lattice (top), FF square lattice (middle), and FF honeycomb lattice (bottom).}
\label{fig:ChainSize}
\end{figure}

The average number of dual edges that the GLM algorithm encounters in each successful attempt of the GLM (which we will call {\it chain} size), and the average number of dual edges that form the partition boundary (which we will call {\it loop} size), provide a measure of the amount of work that the GLM algorithm needs to perform in each iteration. How these two values change with temperature and system size is of interest in determining the efficiency of the algorithm. As illustrated in Fig.~\ref{fig:LoopSize}, for unperturbed FF Ising models, the loop size is independent of lattice size, and increases slightly as temperature decreases. As illustrated in Fig.~\ref{fig:ChainSize}, for unperturbed FF Ising models, the chain size also appears to increase as temperature decreases. Chain size tends to increase with lattice size, however the rate of increase decreases with lattice size, so that the chain size appears to approach a limit as lattice size increases. The increase in chain size with lattice size for smaller systems is likely due to a finite-size effect, meaning that for large lattices both chain size and loop size will saturate.  Therefore, the work performed in each GLM attempt is proportional to a fixed multiple of the work performed in each SSF on a single spin. This computational complexity of the GLM attempt is the motivation behind each GLM iteration consisting of one SSF sweep followed by a $N/30$ GLM attempts \footnote{This number was found to provide a good balance between SSF and GLM updates on the models studied.} in this paper. In this way, the work performed in each GLM iteration is expected to be comparable to a fixed multiple of the work performed in a SSF sweep. More rigorous study is required to determine whether the computational complexity of each GLM iteration is a multiple of the work performed in a SSF sweep.

\subsection{Results on perturbed models} \label{pert}

\begin{figure}
\includegraphics[width=6.5cm]{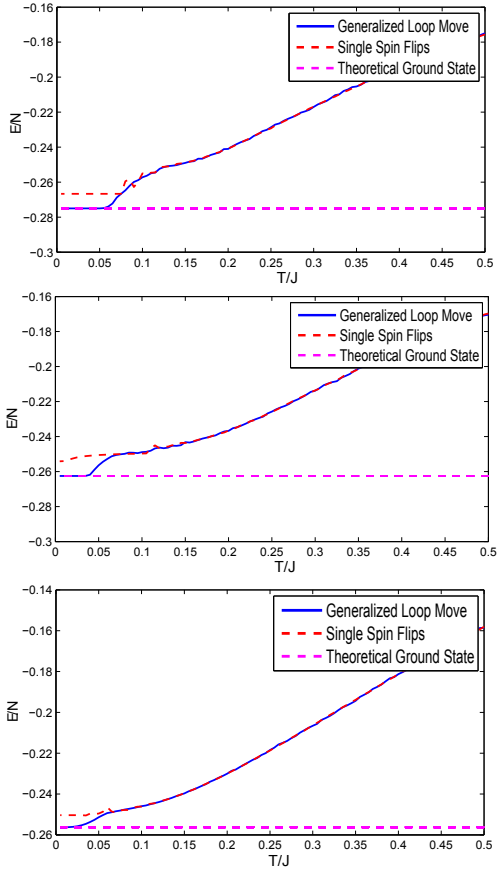}
\caption{Energy per Spin for the Perturbed AFM triangular lattice (top), Perturbed FF square lattice (middle), and Perturbed FF honeycomb lattice (bottom).}
\label{fig:PerturbedAnnealingEnergy}
\end{figure}

We now explore the effect of anisotropy in the Ising interaction on the performance of the MC algorithm, with SSF and GLM updates.  In the following, we define perturbed Ising models by weakening one bond per plaquette in Eq.~(\ref{HAM}),
where two different strengths of $J_{ij}$ are used: $J$ and $J'$. The location of weakened bonds are defined via the geometry illustrated in
 Fig.~\ref{fig:Perturbed Fully Frustrated Ising Models}). The one weakened bond per plaquette has the strength $J'/J = 0.9$.
In the AFM triangular case, $\delta_{ij}=1$ for all bonds, and weakened bonds are chosen arbitrarily to occur along the rows of the lattice; in the two FF models, the weakened bond is chosen to correspond to the FM bond with $\delta_{ij} = -1$ for simplicity.
In each case, the degenerate manifold of groundstates is lifted resulting in two unique groundstates (related by spin inversion symmetry) where the unsatisfied bonds are placed uniquely at the perturbed bonds. This perturbation presents a greater difficulty for SSF MC algorithms to find the true groundstate.  As illustrated in Fig.~\ref{fig:PerturbedAnnealingEnergy}, SSF updates are unable to find the true groundstate energy in any of the three perturbed Ising models through an annealing procedure. GLM updates on the other hand are able to find the true groundstate through annealing (as well as in quenched simulations ran at a single $T \ll J$).   It is interesting to note that at low temperatures, the internal energy curve for GLM is noticeably smoother than the internal energy curve for SSF alone. This suggests that GLM is able to converge to the equilibrium faster than SSF alone.

\begin{figure}
\includegraphics[width=5.5cm]{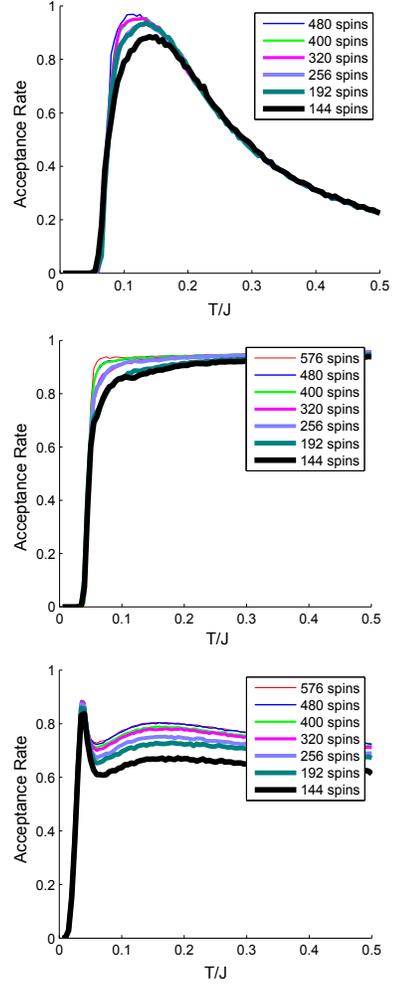}
\caption{Acceptance Rate of GLM on the Perturbed AFM triangular lattice (top), Perturbed FF square lattice (middle), and Perturbed FF honeycomb lattice (bottom).}
\label{fig:PerturbedAcceptanceRate}
\end{figure}

As can be seen in Figure \ref{fig:PerturbedAcceptanceRate}, for the perturbed AFM triangular lattice model, the acceptance rate of GLM updates increases as the temperature decreases between $T = 0.5$ and 0.1, and then sharply drops off to 0 below $T = 0.1$.  For the perturbed fully frustrated square and honeycomb lattices, the acceptance rate is moderately high and stable between $T = 0.5$ and $T = 0.05$ and sharply drops to 0 below $T = 0.05$.  This abrupt drop to zero corresponds to the finite-$T$ phase transition expected to occur in these models at a temperature proportional to the perturbed energy scale, which the GLM updates are able to find cleanly.

\begin{figure}
\includegraphics[width=6.5cm]{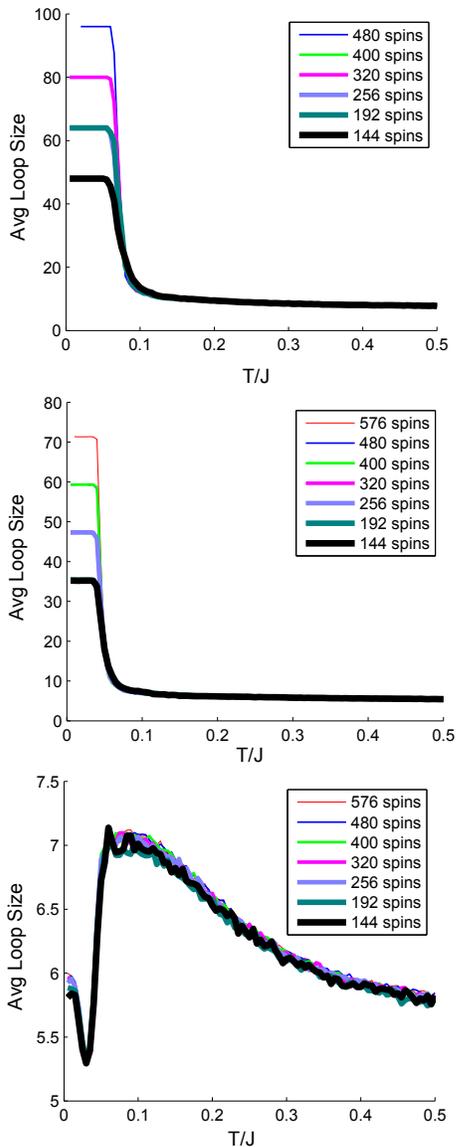}
\caption{Loop Size for the Perturbed AFM triangular lattice (top), Perturbed FF square lattice (middle), and Perturbed FF honeycomb lattice (bottom).}
\label{fig:PerturbedLoopSize}
\end{figure}

\begin{figure}
\includegraphics[width=6.5cm]{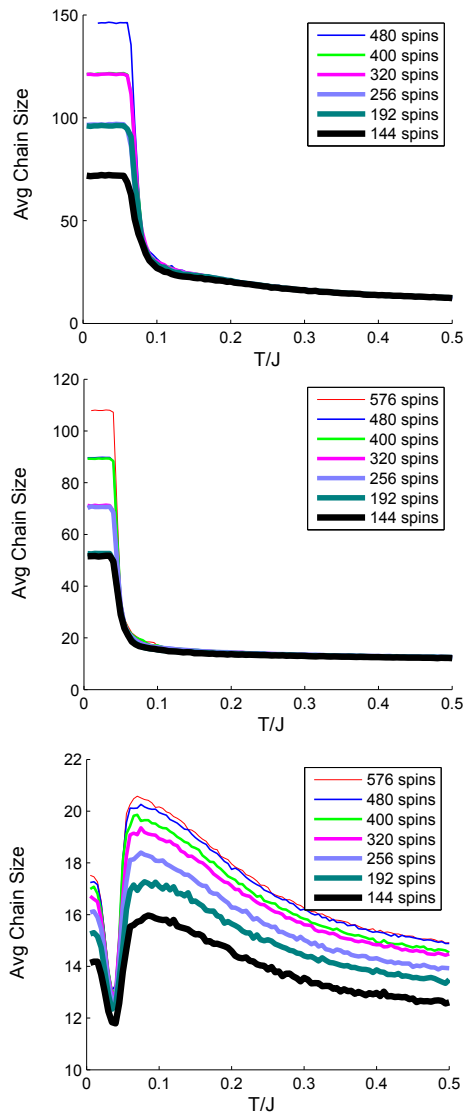}
\caption{Chain Size for the Perturbed AFM triangular lattice (top), Perturbed FF square lattice (middle), and Perturbed FF honeycomb lattice (bottom).}
\label{fig:PerturbedChainSize}
\end{figure}

In the case of perturbed Ising models, temperature has a significant effect on loop and chain size. As illustrated in Figure \ref{fig:PerturbedLoopSize}, for perturbed AFM triangle and FF Square lattices, at high temperatures ($T = 0.5$ to 0.1) loop size appears independent of lattices size except at low temperatures (below the phase transition) 
where loop size appears to grow linearly with the shortest side of the lattice. This is expected based on observing GLM updates in action. For perturbed AFM triangle and FF square lattices at low temperatures, the GLM algorithm tends to encounter spin configurations where the only possible partition boundaries are those that wrap around the lattice. For the perturbed FF honeycomb lattice, loop size appears independent of lattice size regardless of temperature. This is also expected because, based on observation, the GLM algorithm tends to find local partition boundaries that do not grow with lattice size at all temperatures. As illustrated in Figure \ref{fig:PerturbedChainSize}, chain size behaves similar to loop size with respect to lattice size and temperature. These results on loop size and chain size suggest that in the groundstate of some perturbed Ising models (such as AFM triangle and FF square) the amount of work the GLM algorithm needs to perform in each GLM attempt will grow linearly with size of the lattice at low temperatures. This suggests that the amount of work done in each GLM iteration (one SSF iteration followed by $x$ attempts of the GLM where $x=N/30$) will grow quadratically with system size at low temperatures. 
This expensive algorithm cost can be avoided in practical situations by recognizing it as a sign of the underlying long-range order of the system - extensive sampling by the GLM update is not necessary in such a case.  Regardless, further simulation results are required to determine the computational complexity of the GLM update in this interesting case.

\section{Discussion} \label{discuss}

In this paper, we have presented a generalized loop move (GLM) update for a class of frustrated Ising models on two-dimensional lattices which are composed of bond-sharing plaquettes.
The algorithm is designed to allow efficient Metropolis Monte Carlo updates in a degenerate or quasi-degenerate set of spin configurations  
that contribute to an extensive manifold of disordered low-energy states.  We have thoroughly described the algorithm with the goal of making it easy to implement in a wide variety of models, and have provided a rigorous proof of detailed balance in the general case.

We implemented and tested the algorithm on three prototypical lattice models, the antiferromagnetic triangular-lattice Ising model, and the fully-frustrated square- and honeycomb-lattice Ising models.  All three models admit a disordered, extensively-degenerate manifold of configurations in their unperturbed groundstate.  The GLM update is demonstrated to complement traditional single spin-flip updates in allowing the simulation to find all configurations that contribute to the degenerate manifold at low temperatures.  In contrast to single spin-flips, the GLM update is able to sample configurations in the degenerate subspace with a very high acceptance rate, which is independent of temperature for $T \ll J$.  The GLM achieves this while maintaining a very efficient $\mathcal{O}(N)$ complexity for each $N$-site lattice.

Perturbing the models by weakening one bond per plaquette (triangle, square or hexagon) causes them to select a unique spin configuration in their groundstate.  Although single spin-flip updates typically cannot find this unique state due to the large energy barriers associated with sampling the quasi-degenerate manifold of states, we have shown that the GLM updates are capable of finding the equilibrium groundstate in a highly efficient manner.

We expect that our current work in describing and characterizing the GLM updates will lead to their use in other frustrated Ising spin models of physical importance.  The GLM update, as described in this paper, can be straight-forwardly generalized to work on other frustrated Ising models on two-dimensional lattices of bond-sharing plaquettes.  This includes models where the exchange $J_{ij}$ is randomly frustrated, such as Edwards-Anderson spin glass models \cite{Bieche}, which may facilitate the exploration of spin glass phases and phase transitions in a plethora of related models in the future.  Finally, we expect that the GLM moves will be particularly useful in more realistic models of frustrated materials, where additional long-range interactions, local perturbations, or pressure-induced interaction anisotropy leads to the selection of a unique groundstate.

\section{Acknowledgments}

We thank S. Inglis and Geoff Sanders for many useful discussions.
This work was made possible by the facilities of the Shared Hierarchical 
Academic Research Computing Network (SHARCNET:www.sharcnet.ca) and Compute/Calcul Canada.  
Support was provided by NSERC of Canada.

\bibliography{refs}

\begin{thebibliography}{21}
\expandafter\ifx\csname natexlab\endcsname\relax\def\natexlab#1{#1}\fi
\expandafter\ifx\csname bibnamefont\endcsname\relax
  \def\bibnamefont#1{#1}\fi
\expandafter\ifx\csname bibfnamefont\endcsname\relax
  \def\bibfnamefont#1{#1}\fi
\expandafter\ifx\csname citenamefont\endcsname\relax
  \def\citenamefont#1{#1}\fi
\expandafter\ifx\csname url\endcsname\relax
  \def\url#1{\texttt{#1}}\fi
\expandafter\ifx\csname urlprefix\endcsname\relax\def\urlprefix{URL }\fi
\providecommand{\bibinfo}[2]{#2}
\providecommand{\eprint}[2][]{\url{#2}}

\bibitem[{\citenamefont{Metropolis et~al.}(1953)\citenamefont{Metropolis,
  Rosenbluth, Rosenbluth, Teller, and Teller}}]{metrop}
\bibinfo{author}{\bibfnamefont{N.}~\bibnamefont{Metropolis}},
  \bibinfo{author}{\bibfnamefont{A.~W.} \bibnamefont{Rosenbluth}},
  \bibinfo{author}{\bibfnamefont{M.~N.} \bibnamefont{Rosenbluth}},
  \bibinfo{author}{\bibfnamefont{A.~H.} \bibnamefont{Teller}},
  \bibnamefont{and} \bibinfo{author}{\bibfnamefont{E.}~\bibnamefont{Teller}},
  \bibinfo{journal}{J. Chem. Phys.} \textbf{\bibinfo{volume}{21}},
  \bibinfo{pages}{1087} (\bibinfo{year}{1953}).

\bibitem[{\citenamefont{Liu}(2001)}]{Liu}
\bibinfo{author}{\bibfnamefont{J.}~\bibnamefont{Liu}},
  \emph{\bibinfo{title}{Monte Carlo Strategies in Scientiﬁc Computing}}
  (\bibinfo{publisher}{Springer}, \bibinfo{year}{2001}).

\bibitem[{\citenamefont{Moore}(1965)}]{moore}
\bibinfo{author}{\bibfnamefont{G.~E.} \bibnamefont{Moore}},
  \bibinfo{journal}{Electronics} \textbf{\bibinfo{volume}{38}},
  \bibinfo{pages}{8} (\bibinfo{year}{1965}).

\bibitem[{\citenamefont{Swendsen and Wang}(1987)}]{SW}
\bibinfo{author}{\bibfnamefont{R.~H.} \bibnamefont{Swendsen}} \bibnamefont{and}
  \bibinfo{author}{\bibfnamefont{J.-S.} \bibnamefont{Wang}},
  \bibinfo{journal}{Phys. Rev. Lett.} \textbf{\bibinfo{volume}{58}},
  \bibinfo{pages}{86} (\bibinfo{year}{1987}).

\bibitem[{\citenamefont{Wolff}(1989)}]{Wolff}
\bibinfo{author}{\bibfnamefont{U.}~\bibnamefont{Wolff}},
  \bibinfo{journal}{Phys. Rev. Lett.} \textbf{\bibinfo{volume}{62}},
  \bibinfo{pages}{361} (\bibinfo{year}{1989}).

\bibitem[{\citenamefont{Wannier}(1950)}]{Wannier}
\bibinfo{author}{\bibfnamefont{G.~H.} \bibnamefont{Wannier}},
  \bibinfo{journal}{Phys. Rev.} \textbf{\bibinfo{volume}{79}},
  \bibinfo{pages}{357} (\bibinfo{year}{1950}).

\bibitem[{\citenamefont{Houtappel}(1950)}]{Houtappel}
\bibinfo{author}{\bibfnamefont{R.~M.~F.} \bibnamefont{Houtappel}},
  \bibinfo{journal}{Physica} \textbf{\bibinfo{volume}{16}},
  \bibinfo{pages}{425} (\bibinfo{year}{1950}).

\bibitem[{\citenamefont{Barkema and Newman}(1998)}]{BN}
\bibinfo{author}{\bibfnamefont{G.~T.} \bibnamefont{Barkema}} \bibnamefont{and}
  \bibinfo{author}{\bibfnamefont{M.~E.~J.} \bibnamefont{Newman}},
  \bibinfo{journal}{Phys. Rev. E} \textbf{\bibinfo{volume}{57}},
  \bibinfo{pages}{1155} (\bibinfo{year}{1998}).

\bibitem[{\citenamefont{Bramwell and Gingras}(2001)}]{SpinIce}
\bibinfo{author}{\bibfnamefont{S.~T.} \bibnamefont{Bramwell}} \bibnamefont{and}
  \bibinfo{author}{\bibfnamefont{M.~J.~P.} \bibnamefont{Gingras}},
  \bibinfo{journal}{Science} \textbf{\bibinfo{volume}{294}},
  \bibinfo{pages}{1495} (\bibinfo{year}{2001}).

\bibitem[{\citenamefont{Melko et~al.}(2001)\citenamefont{Melko, den Hertog, and
  Gingras}}]{LRO}
\bibinfo{author}{\bibfnamefont{R.~G.} \bibnamefont{Melko}},
  \bibinfo{author}{\bibfnamefont{B.~C.} \bibnamefont{den Hertog}},
  \bibnamefont{and} \bibinfo{author}{\bibfnamefont{M.~J.~P.}
  \bibnamefont{Gingras}}, \bibinfo{journal}{Phys. Rev. Lett.}
  \textbf{\bibinfo{volume}{87}}, \bibinfo{pages}{067203}
  (\bibinfo{year}{2001}).

\bibitem[{\citenamefont{Shinaoka and Motome}(2010)}]{loops2}
\bibinfo{author}{\bibfnamefont{H.}~\bibnamefont{Shinaoka}} \bibnamefont{and}
  \bibinfo{author}{\bibfnamefont{Y.}~\bibnamefont{Motome}}
  (\bibinfo{year}{2010}), \eprint{arXiv:1006.4300}.

\bibitem[{\citenamefont{Orend\'a\ifmmode~\check{c}\else \v{c}\fi{}
  et~al.}(2007)\citenamefont{Orend\'a\ifmmode~\check{c}\else \v{c}\fi{}, Hanko,
  \ifmmode \check{C}\else \v{C}\fi{}i\ifmmode~\check{z}\else \v{z}\fi{}m\'ar,
  Orend\'a\ifmmode~\check{c}\else \v{c}\fi{}ov\'a, Shirai, and
  Bramwell}}]{loops1}
\bibinfo{author}{\bibfnamefont{M.}~\bibnamefont{Orend\'a\ifmmode~\check{c}\else
  \v{c}\fi{}}}, \bibinfo{author}{\bibfnamefont{J.}~\bibnamefont{Hanko}},
  \bibinfo{author}{\bibfnamefont{E.}~\bibnamefont{\ifmmode \check{C}\else
  \v{C}\fi{}i\ifmmode~\check{z}\else \v{z}\fi{}m\'ar}},
  \bibinfo{author}{\bibfnamefont{A.}~\bibnamefont{Orend\'a\ifmmode~\check{c}\e%
lse \v{c}\fi{}ov\'a}},
  \bibinfo{author}{\bibfnamefont{M.}~\bibnamefont{Shirai}}, \bibnamefont{and}
  \bibinfo{author}{\bibfnamefont{S.~T.} \bibnamefont{Bramwell}},
  \bibinfo{journal}{Phys. Rev. B} \textbf{\bibinfo{volume}{75}},
  \bibinfo{pages}{104425} (\bibinfo{year}{2007}).

\bibitem[{\citenamefont{Jaubert et~al.}(2008)\citenamefont{Jaubert, Chalker,
  Holdsworth, and Moessner}}]{PeterKast}
\bibinfo{author}{\bibfnamefont{L.~D.~C.} \bibnamefont{Jaubert}},
  \bibinfo{author}{\bibfnamefont{J.~T.} \bibnamefont{Chalker}},
  \bibinfo{author}{\bibfnamefont{P.~C.~W.} \bibnamefont{Holdsworth}},
  \bibnamefont{and} \bibinfo{author}{\bibfnamefont{R.}~\bibnamefont{Moessner}},
  \bibinfo{journal}{Phys. Rev. Lett.} \textbf{\bibinfo{volume}{100}},
  \bibinfo{pages}{067207} (\bibinfo{year}{2008}).

\bibitem[{\citenamefont{Jaubert et~al.}(2010)\citenamefont{Jaubert, Chalker,
  Holdsworth, and Moessner}}]{Topo}
\bibinfo{author}{\bibfnamefont{L.~D.} \bibnamefont{Jaubert}},
  \bibinfo{author}{\bibfnamefont{J.}~\bibnamefont{Chalker}},
  \bibinfo{author}{\bibfnamefont{P.~C.} \bibnamefont{Holdsworth}},
  \bibnamefont{and} \bibinfo{author}{\bibfnamefont{R.}~\bibnamefont{Moessner}},
  \bibinfo{journal}{arXiv:1003.4896}  (\bibinfo{year}{2010}).

\bibitem[{\citenamefont{Andrews et~al.}(2009)\citenamefont{Andrews, De~Sterck,
  Inglis, and Melko}}]{Shawn}
\bibinfo{author}{\bibfnamefont{S.}~\bibnamefont{Andrews}},
  \bibinfo{author}{\bibfnamefont{H.}~\bibnamefont{De~Sterck}},
  \bibinfo{author}{\bibfnamefont{S.}~\bibnamefont{Inglis}}, \bibnamefont{and}
  \bibinfo{author}{\bibfnamefont{R.~G.} \bibnamefont{Melko}},
  \bibinfo{journal}{Phys. Rev. E} \textbf{\bibinfo{volume}{79}},
  \bibinfo{pages}{041127} (\bibinfo{year}{2009}).

\bibitem[{\citenamefont{Moessner and Sondhi}(2001)}]{Moessner}
\bibinfo{author}{\bibfnamefont{R.}~\bibnamefont{Moessner}} \bibnamefont{and}
  \bibinfo{author}{\bibfnamefont{S.~L.} \bibnamefont{Sondhi}},
  \bibinfo{journal}{Phys. Rev. B} \textbf{\bibinfo{volume}{63}},
  \bibinfo{pages}{224401} (\bibinfo{year}{2001}).

\bibitem[{\citenamefont{Fisher}(1961)}]{square1}
\bibinfo{author}{\bibfnamefont{M.~E.} \bibnamefont{Fisher}},
  \bibinfo{journal}{Phys. Rev.} \textbf{\bibinfo{volume}{124}},
  \bibinfo{pages}{1664} (\bibinfo{year}{1961}).

\bibitem[{\citenamefont{Wolff and Zittartz}(1982)}]{WolffZ}
\bibinfo{author}{\bibfnamefont{W.~F.} \bibnamefont{Wolff}} \bibnamefont{and}
  \bibinfo{author}{\bibfnamefont{J.}~\bibnamefont{Zittartz}},
  \bibinfo{journal}{Z. Phys. B} \textbf{\bibinfo{volume}{49}},
  \bibinfo{pages}{129} (\bibinfo{year}{1982}).

\bibitem[{\citenamefont{Bieche et~al.}(1980)\citenamefont{Bieche, Maynard,
  Rammal, and Uhry}}]{Bieche}
\bibinfo{author}{\bibfnamefont{I.}~\bibnamefont{Bieche}},
  \bibinfo{author}{\bibfnamefont{R.}~\bibnamefont{Maynard}},
  \bibinfo{author}{\bibfnamefont{R.}~\bibnamefont{Rammal}}, \bibnamefont{and}
  \bibinfo{author}{\bibfnamefont{J.~P.} \bibnamefont{Uhry}},
  \bibinfo{journal}{Journal of Physics A} \textbf{\bibinfo{volume}{13}},
  \bibinfo{pages}{2553} (\bibinfo{year}{1980}).

\bibitem[{\citenamefont{Edmonds}(1965{\natexlab{a}})}]{edmonds1}
\bibinfo{author}{\bibfnamefont{J.}~\bibnamefont{Edmonds}},
  \bibinfo{journal}{Can. J. Math.} \textbf{\bibinfo{volume}{17}},
  \bibinfo{pages}{449} (\bibinfo{year}{1965}{\natexlab{a}}).

\bibitem[{\citenamefont{Edmonds}(1965{\natexlab{b}})}]{edmonds2}
\bibinfo{author}{\bibfnamefont{J.}~\bibnamefont{Edmonds}}, \bibinfo{journal}{J.
  Res. NBS.} \textbf{\bibinfo{volume}{69B}}, \bibinfo{pages}{125}
  (\bibinfo{year}{1965}{\natexlab{b}}).

\end{thebibliography}

\appendix
\section{Proof of Detailed Balance}  \label{appendixA}

To prove detailed balance, we need to show that given any two spin
configurations $S_{1}$ and $S_{2}$, the following equation is satisfied:\begin{equation}
\Pi(S_{1})T(S_{1}\rightarrow S_{2})=\Pi(S_{2})T(S_{2}\rightarrow S_{1})\end{equation}
where $\Pi(S)$ denotes the Boltzmann probability of spin configuration
$S$, and $T(S_{i}\rightarrow S_{j})$ denote the transition probability
in the Generalized Loop Move algorithm from spin configuration $S_{i}$
to spin configuration $S_{j}$.

\subsection{Preliminary notations and definitions:}

For ease of exposition, it is convenient to establish the following
notations and definitions:
\begin{itemize}
\item Let $\Gamma$ be the set of all possible dual edge chains that the
Loop Generating subroutine can generate (i.e. $\Gamma$ is the set
of dual edge chains that are alternatingly dual to satisfied and unsatisfied
bonds and end in a simple cycle).
\item Take any 2 spin configurations $S_{1}$ and $S_{2}$, and let:

\begin{itemize}
\item $\Gamma_{1}=\{\gamma\in\Gamma|\:\gamma$ ends in a closed loop and
flipping all the spins inside the closed loop takes the spin system
from $\ensuremath{S_{1}}$ to $\ensuremath{S_{2}\}}$
\item $\Omega_{1}=\{$$(\gamma_{1},\gamma_{2})\in(\Gamma\times\Gamma)|$$\gamma_{1},\gamma_{2}$
ends in cuts $L_{1},\: L_{2}$ and flipping all the spins inside $L_{1}\cup L_{2}$
takes the spin system from $S_{1}$ to $S_{2}$\}
\item $\Psi_{1}=\Gamma_{1}\cup\Omega_{1}$
\item $\Gamma_{2}=\{\gamma\in\Gamma|\:\gamma$ end in a closed loop and
flipping all the spins inside the closed loop takes the spin system
from $\ensuremath{S_{2}}$ to $\ensuremath{S_{1}\}}$
\item $\Omega_{2}=\{$$(\gamma_{1},\gamma_{2})\in(\Gamma\times\Gamma)|$$\gamma_{1},\gamma_{2}$
ends in cuts $L_{1},\: L_{2}$ and flipping all the spins inside $L_{1}\cup L_{2}$
takes the spin system from $S_{2}$ to $S_{1}$\}
\item $\Psi_{2}=\Gamma_{2}\cup\Omega_{2}$
\item $\Psi_{1}^{'}=\{\psi\in\Psi_{1}|\: P_{accept}(\psi,\: S_{1})>0\}$
where $P_{accept}$ is defined as in Equation \ref{eq:Paccept 1}
or \ref{eq:Paccept2}
\item $\Psi_{2}^{'}=\{\psi\in\Psi_{2}|\: P_{accept}(\psi,\: S_{2})>0\}$
where $P_{accept}$ is defined as in Equation \ref{eq:Paccept 1}
or \ref{eq:Paccept2}
\item $\Psi_{1}^{'}$ and $\Psi_{2}^{'}$ are defined as above to facilitate the proof of detailed balance.
\item define $F$: $\Gamma\cup(\Gamma\times\Gamma)\rightarrow\Gamma\cup(\Gamma\times\Gamma)$
such that $F(\gamma)=f(\gamma)\;\forall\gamma\in\Gamma$ and $F((\gamma_{1},\gamma_{2}))=(f(\gamma_{1}),\: f(\gamma_{2}))\;\forall(\gamma_{1},\gamma_{2})\in(\Gamma\times\Gamma)$ 
\item Let $\tilde{F}$ be the restriction of $F$ to $\Psi_{1}^{'}$
\end{itemize}
For convenience, we will also define $P_{select}((\gamma_{1},\gamma_{2}),\: S)=P_{select}(\gamma_{1},\: S)P_{select}(\gamma_{2},\: S)\;\forall(\gamma_{1},\gamma_{2})\in(\Gamma\times\Gamma)$

Note: $T(S_{i}\rightarrow S_{j})=\sum_{\psi\in\Psi_{i}^{'}}P_{select}(\psi,\: S_{i})P_{accept}(\psi,\: S_{i})$,
where $(i,\: j)=(1,\:2)$ or $(2,\:1)$

\end{itemize}

\subsection{Proof of Detailed Balance}

If we can prove the following 3 subclaims:
\begin{enumerate}
\item $\tilde{F}(\Psi_{1}^{'})\subseteq\Psi_{2}^{'}$ (this is used to prove
subclaim 2).
\item $\tilde{F}$ is bijective between $\Psi_{1}^{'}$ and $\Psi_{2}^{'}$.
\item $\forall\psi\in\Psi^{'}_{1},$ $\Pi(S_{1})P_{select}(\psi,\: S_{1})P_{accept}(\psi,\: S_{1})=\Pi(S_{2})P_{select}(\tilde{F}(\psi),\: S_{2})P_{accept}(\tilde{F}(\psi),\: S_{2})$.
\end{enumerate}
\begin{widetext}
Then, the proof for detailed balance follows naturally: \begin{eqnarray*}
 & {\displaystyle \Sigma_{\psi\in\Psi_{1}^{'}}\Pi(S_{1})P_{select}(\psi,\: S_{1})P_{accept}(\psi,\: S_{1})=\Sigma_{\psi\in\Psi_{1}^{'}}\Pi(S_{2})P_{select}(\tilde{F}(\psi),\: S_{2})P_{accept}(\tilde{F}(\psi),\: S_{2})} & \text{}\\
 & \text{by subclaim 3}\\
\Rightarrow & {\displaystyle \Sigma_{\psi\in\Psi_{1}^{'}}\Pi(S_{1})P_{select}(\psi,\: S_{1})P_{accept}(\psi,\: S_{1})=\Sigma_{\delta\in\Psi_{2}^{'}}\Pi(S_{2})P_{select}(\delta,\: S_{2})P_{accept}(\delta,\: S_{2})} & \text{ }\\
 & \text{because \ensuremath{\tilde{F}}is bijective between \ensuremath{\Psi_{1}^{'}}and \ensuremath{\Psi_{2}^{'}}}\\
\Rightarrow & {\displaystyle \Pi(S_{1})\Sigma_{\psi\in\Psi_{1}^{'}}P_{select}(\psi,\: S_{1})P_{accept}(\psi,\: S_{1})=\Pi(S_{2})\Sigma_{\delta\in\Psi_{2}^{'}}P_{select}(\delta,\: S_{2})P_{accept}(\delta,\: S_{2})}\\
\Rightarrow & \Pi(S_{1})T(S_{1}\rightarrow S_{2})=\Pi(S_{2})T(S_{2}\rightarrow S_{1})\end{eqnarray*}

\subsection{Proof of subclaims}

\subsubsection*{Subclaim 1: $\tilde{F}(\Psi_{1}^{'})\subseteq\Psi_{2}^{'}$}

Take any $\psi\in\Psi_{1}^{'}$, show that $\tilde{F}(\psi)\in\Psi_{2}^{'}$
\begin{enumerate}
\item $\tilde{F}(\psi)\in\Psi_{2}$ because:

\begin{enumerate}
\item $\psi\in\Psi_{1}$ implies $\psi$ ends in a closed loop or union
of complementary cuts that forms a partition boundary around a group
of spins that if flipped will take the spin configuration from $S_{1}$
to $S_{2}$. Therefore, $\tilde{F}(\psi)$ ends in a closed loop or
union of complementary cuts that forms a partition boundary around
the same group of spins, which if flipped will take the spin configuration
from $S_{2}$ back to $S_{1}$
\item and $P_{select}(\tilde{F}(\psi),\; S_{2})>0$ because \begin{eqnarray*}
\psi\epsilon\in\Psi_{1}^{'} & \Rightarrow & P_{accept}(\psi,\: S_{1})>0\\
 & \Rightarrow & min(1,\; e^{-\beta[E(S_{2})-E(S_{1})]}\frac{P_{select}(\tilde{F}(\psi),\; S_{2})}{P_{select}(\psi,\; S_{1})})>0\\
 & \Rightarrow & P_{select}(\tilde{F}(\psi),\; S_{2})>0\end{eqnarray*}

\end{enumerate}
\item \[P_{accept}(\tilde{F}(\psi),\; S_{2})=min(1,\; e^{-\beta[E(S_{2})-E(S_{1})]}\frac{P_{select}(\psi,\: S_{1})}{P_{select}(\tilde{F}(\psi),\: S_{2})})>0  \]
because $\psi\in\Psi_{1}^{'}\subseteq\Psi_{1}\Rightarrow P_{select}(\psi,\: S_{1})>0$.
\item 1 \& 2 $\Longrightarrow\tilde{F}(\psi)\in\Psi_{2}^{'}$. Therefore,
$\tilde{F}(\Psi_{1}^{'})\subseteq\Psi_{2}^{'}$
\end{enumerate}

\subsubsection*{Subclaim 2: $\tilde{F}$ is bijective between $\Psi_{1}^{'}$ and $\Psi_{2}^{'}$}
\begin{enumerate}
\item $\tilde{F}$ is injective because $\tilde{F}(\Psi_{1}^{'})\subseteq\Psi_{2}^{'}$
by subclaim 1 and take any $\psi_{1}$, $\psi_{2}$ $\in\Psi_{1}$,
$\tilde{F}(\psi_{1})=\tilde{F}(\psi_{2})\Rightarrow\psi_{1}=\psi_{2}$ by definition of $f()$
\item $\tilde{F}$ is subjective because for any $\delta\in\Psi_{2}^{'}$,
$F(\delta)\in\Psi_{1}^{'}$ by subclaim 1, and $\tilde{F}(F(\delta))=\delta$.
\end{enumerate}

\subsubsection*{Subclaim 3: For any $\psi\in\Psi^{'}_{1}$, show that $\Pi(S_{1})P_{select}(\psi,\: S_{1})P_{accept}(\psi,\: S_{1})=\Pi(S_{2})P_{select}(\tilde{F}(\psi),\: S_{2})P_{accept}(\tilde{F}(\psi),\: S_{2})$}

Let $a=\Pi(S_{1})P_{select}(\psi,\: S_{1})$, and $b=\Pi(S_{2})P_{select}(\tilde{F}(\psi),\: S_{2})$.
Then,\begin{eqnarray*}
\Pi(S_{1})P_{select}(\psi,\: S_{1})P_{accept}(\psi,\: S_{1}) & = & \Pi(S_{1})P_{select}(\psi,\: S_{1})\min(1,\frac{\Pi(S_{2})}{\Pi(S_{1})}\frac{P_{select}(\tilde{F}(\psi),\: S_{2})}{P_{select}(\psi,\: S_{1})})\\
 & = & a\: \min(1,\:\frac{b}{a})\\
 & = & \min(a,\: b)\\
\Pi(S_{2})P_{select}(\tilde{F}(\psi),\: S_{2})P_{accept}(\tilde{F}(\psi),\: S_{2}) & = & \Pi(S_{2})P_{select}(\tilde{F(}\psi),\: S_{2})\min(1,\frac{\Pi(S_{1})}{\Pi(S_{2})}\frac{P_{select}(\psi,\: S_{1})}{P_{select}(\tilde{F}(\psi),\: S_{2})})\\
 & = & b\: \min(1,\:\frac{a}{b})\\
 & = & \min(a,\: b)\end{eqnarray*}
Therefore $\Pi(S_{1})P_{select}(\psi,\: S_{1})P_{accept}(\psi,\: S_{1})=\Pi(S_{2})P_{select}(\tilde{F}(\psi),\: S_{2})P_{accept}(\tilde{F}(\psi),\: S_{2})$

\end{widetext}

\end{document}